\documentstyle[prd,aps]{revtex}

 \mathchardef\ScriptA="7241
 \mathchardef\ScriptB="7242 
 \mathchardef\ScriptC="7243
 \mathchardef\ScriptD="7244
 \mathchardef\ScriptE="7245
 \mathchardef\ScriptF="7246
 \mathchardef\ScriptG="7247
 \mathchardef\ScriptH="7248
 \mathchardef\ScriptI="7249
 \mathchardef\ScriptJ="724A
 \mathchardef\ScriptK="724B
 \mathchardef\ScriptL="724C
 \mathchardef\ScriptM="724D
 \mathchardef\ScriptN="724E
 \mathchardef\ScriptO="724F
 \mathchardef\ScriptP="7250
 \mathchardef\ScriptQ="7251
 \mathchardef\ScriptR="7252
 \mathchardef\ScriptS="7253
 \mathchardef\ScriptT="7254
 \mathchardef\ScriptU="7255
 \mathchardef\ScriptV="7256
 \mathchardef\ScriptW="7257
 \mathchardef\ScriptX="7258
 \mathchardef\ScriptY="7259
 \mathchardef\ScriptZ="725A

 \mathchardef\#="0023
 \mathchardef\$="0024
 \mathchardef\%="0025
 \mathchardef\ddash="705C
 
 \mathchardef\lwavy="336E
 \mathchardef\rwavy="336F
 \mathchardef\biglwavy="331A
 \mathchardef\bigrwavy="331B
 \mathchardef\bigglwavy="3328
 \mathchardef\biggrwavy="3329
 \mathchardef\littlesum="0350

\tighten
\draft
\begin{document} 
\bibliographystyle{prsty}

\title{
Radiatively Generated Neutrino Masses\\
in $SU(3)_L\times U(1)_N$ Gauge Models
}

\author{
Yutaka Okamoto$^{a}$
\footnote{E-mail:8jspd003@keyaki.cc.u-tokai.ac.jp}
and Masaki Yasu${\grave {\rm e}}^{a,b}$
\footnote{E-mail:yasue@keyaki.cc.u-tokai.ac.jp}
}

\address{\vspace{5mm}$^{a}${\sl Department of Physics, Tokai University}\\
{\sl 1117 KitaKaname, Hiratsuka, Kanagawa 259-1292, Japan}}
\address{\vspace{2mm}$^{b}${\sl Department of Natural Science\\School of Marine
Science and Technology, Tokai University}\\
{\sl 3-20-1 Orido, Shimizu, Shizuoka 424-8610, Japan}}
\date{TOKAI-HEP/TH-9902, June, 1999}
\maketitle

\begin{abstract}
In $SU(3)_L\times U(1)_N$ gauge models for electroweak interactions, we discuss how to 
implement a radiative mechanism of generating Majorana neutrino masses by 
considering the property that the Higgs scalar, which has a coupling to a charged 
lepton ($\ell$)-neutrino ($\nu$) pair, is naturally included.  
The mechanism is shown to work in models with a heavy charged lepton, 
$\omega^+$, in a lepton triplet ($\nu$, $\ell$, $\omega^+$) and with a heavy neutral lepton, 
$N$, in ($\nu$, $\ell$, $N$).  A minimal model with $\ell$ and $\ell^c$ in ($\nu$, $\ell$, $\ell^c$) 
together with a sextet Higgs scalar suffers from a fine-tuning problem to suppress tree-level 
neutrino masses.
\end{abstract}
\vspace{2mm}
It is our common understanding that neutrinos are massive and oscillate among families\cite{EarlyMassive}. 
The first experimental evidence is provided by the SuperKamiokande collaboration\cite{SuperKamiokande}, 
which has observed atmospheric neutrino oscillations.  There have been indications of solar 
neutrino oscillations\cite{SolarDeficit},  which are awaiting to be confirmed.  Neutrinos produced 
at Los Alamos may also oscillate\cite{LSND} and will be tested for their behavior at other 
laboratories\cite{LaboNu}. Theoretically, neutrinos can acquire 
Majorana masses if the lepton number conservation is not respected.  The origin of the lepton number 
violation is currently ascribed either to the formation of Majorana masses for right-handed neutrinos 
or to the existence of interactions involving a charged lepton-neutrino pair. The former is usually 
referred to as a seesaw mechanism\cite{SeeSaw}  and the latter is referred to as a radiative 
mechanism\cite{Radiative}.  Any viable models describing physics beyond the one based on the 
standard $SU(3)_c$ $\times$ $SU(2)_L$ $\times$ $U(1)_Y$ gauge model should contain mechanisms 
that accommodate massive neutrinos.  

Among the possible extensions of the standard model, electroweak models based on 
an $SU(3)_L\times U(1)_N$ gauge group
\cite{SU3U1,RhandedNU,Sextet,RadiativeNuDirac,RadiativeNuSU3,MajoranaNuSextet,GIMinSU3,SUSYSU3U1,SU3U1Application,SU3U1Muonium} 
provide the intriguing aspect that the models predict 
three families of quarks and leptons if the anomaly free condition on $SU(3)_L\times U(1)_N$ and 
the asymptotic free condition on $SU(3)_c$ are imposed.  Another virtue of the models 
lies in the fact that the notion of the lepton number loses its meaning since a lepton triplet 
simultaneously contains the lepton and its antiparticle.  Therefore, interactions that provide a coupling to 
a charged lepton-neutrino pair are expected to be naturally incorporated and can be utilized 
to radiatively generate neutrino masses\cite{RadiativeNuDirac,RadiativeNuSU3}.  The experimentally 
suggested magnitude of neutrino masses is at most of order 1 eV\cite{MassNu}, which is much smaller than 
the mass of the electron.  In the radiative mechanism, the lightness of neutrinos will be 
ensured by the smallness of  the charged lepton masses and by weak couplings associated with 
lepton-number violating interactions. 

In this report, we discuss how lighter masses of neutrinos are generated in $SU(3)_L\times U(1)_N$ 
gauge models.  The minimal model contains left ($L$)-handed lepton states of the neutrino ($\nu_L$) 
and the charged lepton ($\ell_L$) in each family, which belong to the same triplet:
\begin{eqnarray}\label{Eq:LeptonTriplets}
& \psi^{i=1,2,3}_L = \left( \nu^i, \ell^i, \ell^{ci}\right)^T_L:({\bf 3}, 0),
\end{eqnarray}
where $i$ = 1,2,3 stands for the family index, the superscript $c$ of $\ell$ denotes the charge conjugation
 and the values in the parentheses specify quantum numbers 
based on the  ($SU(3)_L$, $U(1)_N$) - symmetry. Let $N/2$ be the $U(1)_N$ quantum number.  Then, the 
hypercharge, $Y$, is given by $Y$ = $-\sqrt{3}\lambda^8$ + $N$ and the electric charge, $Q_{em}$, 
is given by $Q_{em}$ = ($\lambda^3$ + $Y$)/2, where $\lambda^a$ is the $SU(3)$ generator with 
Tr($\lambda^a\lambda^b$) = 2$\delta^{ab}$ ($a,b$ = 1$\sim$8).  The quark sector is described by 
two quark antitriplets and one quark triplet, which are denoted by
\begin{eqnarray}\label{Eq:QuarkTriplets}
& Q^{i=1,2}_L  =  \left( d^i,  -u^i,  J^i\right)^T_L:({\bf 3^\ast}, -\frac{1}{3}),  \ \ \ 
Q^3_L  =  \left( u^3,  d^3,  J^3\right)^T_L:({\bf 3}, \frac{2}{3}), 
\end{eqnarray}
where, for instance,  $Q^3_L$ can be taken as quarks in the third family.  It is obvious that the pure 
$SU(3)_L$ anomaly vanishes since there is an equal number of  triplets of quarks and leptons and 
antitriplets of quarks.  
The right ($R$)-handed partners are classified as
\begin{eqnarray}\label{Eq:ExtraQuarkTriplets}
& u^{i=1,2,3}_R:({\bf 1}, \frac{2}{3}), \ \ \  d^{i=1,2,3}_R:({\bf 1}, -\frac{1}{3}),  \ \ \ 
J^{i=1,2}_R:({\bf 1}, -\frac{4}{3}), \ \ \ J^3_R:({\bf 1}, \frac{5}{3}).  
\end{eqnarray}

The Higgs scalars are used to break $SU(3)_L$ $\times$ $U(1)_N$ down to $U(1)_{em}$ 
via the standard $SU(2)_L$ $\times$ $U(1)_Y$ gauge group, where $U(1)_{em}$ stands for the 
electromagnetic $U(1)$ gauge group.  The minimal set is supplied by 
\begin{eqnarray}\label{Eq:HiggsEtaRhoChi}
& \eta  =  \left( \eta^0,  \eta^-,  \eta^+\right)^T:({\bf 3}, 0),  \
\rho  =  \left( \rho^+,  \rho^0,  \rho^{++}\right)^T:({\bf 3}, 1),  \ 
\chi  =  \left( \chi^-,  \chi^{--},  \chi^0\right)^T:({\bf 3}, -1),  
\end{eqnarray}
which will develop vacuum expectation values (VEV's):
\begin{equation}\label{Eq:VEVs}
\langle 0|\eta |0\rangle = ( v_\eta, 0, 0)^T, \ \    
\langle 0|\rho |0\rangle = (0,  v_\rho, 0)^T, \ \   
\langle 0|\chi |0\rangle = (0,  0, v_\chi)^T.  
\end{equation}
Masses of quarks and leptons are generated by Yukawa interactions:
\begin{eqnarray}\label{Eq:Yukawa}
-{\cal L}_Y  & = &  
\frac{1}{2} \epsilon^{\alpha\beta\gamma} G_{[ij]}{\overline {\left( \psi_{\alpha L}^i \right)^c}}\psi^j_{\beta L}\eta_\gamma
 + \sum_{i=1}^{2}\left( {\overline {Q^i_L}}D^i_R\eta^\ast + {\overline {Q^i_L}}U^i_R\rho^\ast+{\overline {Q^i_L}}C^i_R\chi^\ast \right) 
\nonumber \\
& + &  {\overline {Q^3_L}}U^3_R \eta
 +{\overline {Q^3_L}}D^3_R\rho
 +G_{J3}{\overline {Q^3_L}}J^3_R\chi + ({\rm h.c.})
\end{eqnarray}
with $U^i_R$ = $\sum_{j=1}^3G^i_{uj}u^j_R$, $D^i_R$ = $\sum_{j=1}^3G^i_{dj}d^j_R$ and $C^i_R$ = $\sum_{j=1}^2G^i_{Jj}J^j_R$, 
where $\alpha,\beta,\gamma$ stand for the $SU(3)_L$ indices and  $G$'s ($G_{[ij]}$ = $-G_{[ji]}$) denote 
the Yukawa couplings.  The leptonic part contains lepton-number-violating interactions.  Exotic quarks 
receive their masses through $\langle 0|\chi |0\rangle$.  
Because the lepton mass matrix turns out to be proportional to $G_{[ij]}$, which is antisymmetric 
with respect to the flavor index, it yields one massless lepton and two degenerated 
massive leptons, which are not phenomenologically accepted.  To remedy this situation, 
one can consider three options\cite{ThreeOptions}: 
\begin{enumerate}
\item to introduce a sextet Higgs scalar to produce a flavor-symmetric mass matrix\cite{DiscrteSymmetry}, 
\item to replace $\ell^{ci}_L$ by a heavy charged lepton $\omega^{+i}_L$ with $\omega^{+i}_R$ and $\ell^i_R$  added\cite{HeavyLepton} and 
\item to replace $\ell^{ci}_L$ by the antineutrino $\nu^{ci}_L$ with $\ell^i_R$  added\cite{RhandedNU,RadiativeNuDirac}.
\end{enumerate} 
The rest of our discussions deals with how radiative generation of Majorana neutrino masses is implemented 
in these variants.  In the minimal model, the mass term of $\ell^i_L\ell^{cj}_L$ is always 
accompanied by the lepton-number-violating term of $\nu^i_L\ell^j_L$, 
which is the main source for Majorana neutrino masses.   This property is also inherited by 
the variants. The Higgs scalars exhibit various couplings, which will be 
constrained so as to incorporate a radiative mechanism for neutrino masses.  

Let us first examine the model with a sextet Higgs scalar, $S:({\bf 6^\ast}, 0)$, which gives 
flavor-symmetric Yukawa interactions:
\begin{equation}\label{Eq:YukawaSextet}
G_{\{ ij\}}{\overline {\left( \psi_{\alpha L}^i \right)^c}}\psi^j_{\beta L}S^{\alpha\beta} + {\rm (h.c.)}, 
\end{equation}
where $G_{\{ij\}}$ = $G_{\{ji\}}$.  The $S^{11}$ scalar, $\sigma_1^0$, has a coupling to the neutrino. 
It has been stressed in Ref.\cite{DiscrteSymmetry} that $\sigma_1^0$ 
will readily develop a VEV through other Higgs interactions.  
As a result, neutrinos get Majorana masses at the tree level, which  should be forbidden 
since we are interested in radiative generation of neutrino masses.  Furthermore, since $\sigma_1^0$ 
belongs to a triplet of the standard  $SU(2)_L$ gauge group, phenomenology of weak interactions 
also restricts $\langle 0| \sigma_1^0|0 \rangle$ to be much smaller than the VEV of the $SU(2)_L$-doublet 
Higgs scalar.  One may have a fine-tuning among various Higgs couplings to yield vanishing neutrino 
masses at the tree level and a radiative mechanism is, then, operative\cite{FineTuneNu}.  

We impose an extra symmetry that permits maintaining $\langle 0| \sigma_1^0|0 \rangle$ = 0 
regardless of any Higgs interactions.
\footnote{
There was a suggestion that the condition of 
$\langle 0| \sigma_1^0|0 \rangle$ = 0 could be linked to the masslessness 
of the photon, which, however, may not be the case.  
See Ref.\cite{MajoranaNuSextet} for the details.
}
The discrete symmetry given by 
$\eta$ $\rightarrow$ $-\eta$, $\rho$ $\rightarrow$ $i\rho$, $\chi$ $\rightarrow$ $i\chi$, 
$S$ $\rightarrow$ $-S$, $\psi_L$ $\rightarrow$ $i\psi_L$, $Q^{i=1,2}_L$ $\rightarrow$ $-iQ^i_L$, 
$Q^3_L$ $\rightarrow$ $-Q^3_L$, $u^i_R$ $\rightarrow$ $u^i_R$, $d^i_R$ $\rightarrow$ $id^i_R$, 
$J^{i=1,2}_R$ $\rightarrow$ $J^i_R$ and $J^3_R$ $\rightarrow$ $iJ^3_R$\cite{DiscrteSymmetry} 
forbids dangerous couplings such as
$\eta S\eta$, ${\rm det}S$, 
$\epsilon_{\alpha\beta\gamma}\left( S\eta\right)^\alpha \rho^{c\beta} \chi^{c\gamma}$, 
$\epsilon_{\alpha\beta\gamma}\epsilon_{\alpha'\beta'\gamma'}S^{\alpha\alpha'}S^{\beta\beta'}S^{\gamma\gamma'}$ and 
$\epsilon_{\alpha\beta\gamma}\epsilon_{\alpha'\beta'\gamma'}S^{\alpha\beta}S^{\alpha'\beta'}\rho^{c\gamma}\chi^{c\gamma'}$, 
which tend to align $\langle 0| \sigma_1^0|0 \rangle$ $\neq$ 0\cite{DangerousCoupling}.  
Instead, the required coupling to align $\langle 0| \sigma_1^0|0 \rangle$ = 0 is represented by $\rho S \chi$, 
which is allowed.  Since the Majorana neutrino mass 
term behaves as $\sigma_1^{0\dagger}$, the neutrino ($\nu$) will have an effectively generated 
$\nu\sigma_1^0 \nu$-coupling once it acquires a mass.  This shows that any attempts to 
generate neutrino masses rely upon  
$\langle 0| \sigma_1^0|0 \rangle$ $\neq$ 0.  In other words,  to generate Majorana neutrino masses,  
one needs a Higgs interaction that is not respected by the discrete symmetry. The Higgs potential including 
such an interaction will necessarily yield $\langle 0| \sigma_1^0|0 \rangle$ $\neq$ 0.  Therefore, the 
sextet Higgs scalar should not be employed.  

Next, we examine the second model with heavy charged leptons, which 
are introduced in each family. The lepton sector is modified to:
\begin{eqnarray}\label{Eq:HeavyLepton}
& \psi^i_L = \left( \nu^i, \ell^i, \omega^i\right)^T_L:({\bf 3}, 0), \ \ \ 
\ell^i_R:~ ({\bf 1}, -1), \ \ \ \omega^i_R:~ ({\bf 1}, 1),
\end{eqnarray}
where we have denoted $\omega^{+i}$ by $\omega^i$.  Their Yukawa interactions 
are given by
\begin{equation}\label{Eq:YukawaHeavyLepton}
{\overline {\psi^i_L}}\left( L^i\rho+\Omega^i\chi\right) + {\rm (h.c)}
\end{equation}
with the definitions of 
$L^i$ = $\sum_{j=1}^3G^i_{\ell j}\ell^j_R$ and $\Omega^i$ = $\sum_{j=1}^3G^i_{\omega j}\omega^j_R$.
In this case, the neutrino mass term behaves as 
$\eta^{0\dagger}\eta^{0\dagger}$ instead of $\sigma_1^{0\dagger}$. 
The possible Higgs coupling,
\begin{equation}\label{Eq:HiggsNuHeavyLepton}
\lambda\left( \rho^\dagger\eta \right) \left( \chi^\dagger\eta\right) + ({\rm h.c.}),
\end{equation}
is chosen to be responsible for generating Majorana neutrino masses, where $\lambda$ is 
a coupling strength.  The most general Higgs potential is, thus, described by self-Hermitian terms composed of  
$\phi_\alpha\phi^\dagger_\beta$ ($\phi$ = $\eta$, $\rho$, $\chi$) and by the non-self-Hermitian term of 
$( \rho^\dagger\eta ) ( \chi^\dagger\eta )$ as well as by the familiar term of 
$\epsilon^{\alpha\beta\gamma}\eta_\alpha\rho_\beta\chi_\gamma$\cite{SU3U1}.

The relevant Yukawa interactions, which retain the $\psi\psi\eta$ - part in Eq.(\ref{Eq:Yukawa}) as a 
lepton-number-violating source, are given by
\begin{eqnarray}\label{Eq:NuMasses}
& - & \frac{1}{2}G_{[ij]}\left[
\left( {\overline {\ell^{ci}_R}}\nu^j_L - {\overline {\nu^{ci}_R}}\ell^j_L \right)\eta^+ + 
\left( {\overline {\omega^{ci}_R}}\ell^j_L - {\overline {\ell^{ci}_R}}\omega^j_L \right) \eta^0 + 
\left( {\overline {\nu^{ci}_R}} \omega^j_L - {\overline {\omega^{ci}_R}} \nu^j_L \right) \eta^- 
\right] 
\nonumber \\
&  & \ \ \ \ -G^i_{\ell i}\left( {\overline {\nu^i_L}}\ell^i_R\rho^+ +  {\overline {\ell^i_L}}\ell^i_R\rho^0 \right) 
- G^i_{\omega i}\left( {\overline {\nu^i_L}}\omega^i_R\chi^- +  {\overline {\omega^i_L}}\omega^i_R\chi^0 \right) 
+  ({\rm h.c.}),
\end{eqnarray}
where the charged lepton and heavy lepton mass matrices are assumed to be diagonal for simplicity.   
The one-loop diagrams of the Zee's radiative type\cite{Radiative} contain three processes via 
charged leptons, via heavy charged leptons and via their mixings, which are 
shown in Fig.1(a) $\sim$ (c).  These diagrams correspond to an effective coupling of
\begin{equation}\label{Eq:EffectiveCoupling}
\left(\eta^\dagger\psi_L\right)\epsilon^{\alpha\beta\gamma}\rho_\alpha\chi_\beta\psi_{\gamma L} 
\end{equation}
as a Majorana mass term\cite{Ma}. The resulting neutrino mass matrix turns out to take the following form:
\begin{eqnarray}\label{Eq:NUmass}
\left( \begin{array}{ccc}
	\delta_{11}&	m_{12}+\delta_{12}&	m_{13}+\delta_{13}\\
	m_{12}+\delta_{12}&	\delta_{22}&	m_{23}+\delta_{23}\\
	m_{13}+\delta_{13}&	m_{23}+\delta_{23}&	\delta_{33}\\
\end{array} \right),
\end{eqnarray}
where $m_{ij}$ and $\delta_{ij}$ are calculated to be
\begin{eqnarray}\label{Eq:MatrixEntryIJ}
m_{ij}  &=& \lambda G_{[ij]} 
\left[ 
	\frac{m^2_{\ell_j}-m^2_{\ell_i}}{v^2_\rho}F(m_{\eta^+}^2, m_{\rho^+}^2) 
	+\frac{m^2_{\omega_j}-m^2_{\omega_i}}{v^2_\chi}F(m_{\eta^-}^2, m_{\chi^-}^2) 
\right]  v_\eta v_\rho v_\chi \ \ (i\neq j),\\
\label{Eq:MatrixEntryDeltaIJ}
\delta_{ij}  &=& \frac{\lambda}{4}\sum_{k,\ell}G_{[ik]}G_{[k\ell]}G_{[\ell j]}
\left[ G(m^2_{\omega_k}, m^2_\eta)-G(m^2_{\omega_\ell}, m^2_\eta)\right] v_\eta v_\rho v_\chi \ \ (i\neq j),\\
\label{Eq:MatrixEntryDeltaII}
\delta_{ii}  &=& \frac{\lambda}{4}G_{[12]}G_{[23]}G_{[31]}\sum_{j,k}\epsilon_{ijk}
\left[ G(m^2_{\omega_j}, m^2_\eta)-G(m^2_{\omega_k}, m^2_\eta)\right] v_\eta v_\rho v_\chi,
\end{eqnarray}
and
\begin{equation}\label{Eq:F_x_y}
F(x, y) = \frac{1}{16\pi^2}\frac{\log x-\log y}{x-y},  \ \ \ 
G(x, y) = \frac{1}{16\pi^2}\frac{1}{x-y}\left[ x\frac{\log x-\log y}{x-y}-1 \right]. 
\end{equation}
As mass parameters, $m_{\eta^\pm,\rho^+,\chi^-}$ are the masses of 
Higgs scalars, $\eta^\pm$, $\rho^+$ and $\chi^-$, and $m_{\ell_i}$ ($\equiv$ $G^i_{\ell i}v_\rho$) 
and $m_{\omega_i}$ ($\equiv$ $G^i_{\omega i}v_\chi$) are, respectively, the mass of the $i$-th 
charged lepton and the mass of the $i$-th heavy charged lepton.  The explicit form of $G(x,y)$
 is subject to the condition of  $m_{\eta^+}$ = $m_{\eta^-}$ (=$m_\eta$) with $m_{\ell_i}$ neglected.  
We are anticipating that the mass, $\delta_{ij}$, is extremely small since it is proportional 
to the cubic of the lepton-number-violating coupling of $G_{[ij]}$. 

Expected neutrino phenomenology arising from the mass matrix of (\ref{Eq:NUmass}) 
has been extensively discussed in Ref.\cite{ZeeModelNU}.  
For example, to get bi-maximal neutrino mixing, one has to require that $m_{12}$ $\sim$  $m_{13}$ $\sim$ 
0.01 eV together with $m_{13}$/$m_{23}$ $\sim$ $\Delta m^2_{atm}$/$\Delta m_{\odot}^2$, where 
$\Delta m^2_{atm}$ $\sim$ 10$^{-3}$ eV$^2$ and $\Delta m_{\odot}^2$ $\sim$ 10$^{-5}$ or 10$^{-10}$ 
eV$^2$.  To figure out the order of magnitudes for $G_{[ij]}$, let us specify various mass parameters. 
For values of $v_{\eta,\rho,\chi}$, $v_{\eta,\rho}$ yield masses of weak bosons 
and $v_\chi$ is a source of masses for heavy charged leptons, exotic quarks and exotic gauge bosons. 
So, let us use ($v_\eta$,  $v_\rho$, $v_\chi$) $\sim$ (100, 100, 1000) GeV as typical values, 
$m_{\omega_i}$ $\sim$ 100 GeV as masses for the heavy charged leptons 
and $\ScriptO$(300) GeV as masses for the mediating charged 
Higgs scalars.  We find that, for $\Delta m_{ij}$ representing $\vert m_{\omega_i} - m_{\omega_j} \vert$, 
$m_{ij}$ $\sim$ $G_{[ij]}\lambda (\Delta m_{ij} /1 {\rm GeV})\times 10^{-5}$ (GeV) for the 
heavy-lepton-exchange. For $\Delta m_{ij}$ $\sim$ 10 GeV, where 
the heavy-lepton-exchange dominates the charged-lepton-exchange, we obtain  
$G_{[12]}\lambda$ $\sim$ $G_{[13]}\lambda$ $\sim$ 10$^{-7}$ with 
$G_{[23]}/G_{[13]}$ $\sim$ 10$^{-2}$ or 10$^{-7}$. 

The last example to avoid flavor-antisymmetric mass terms is to put $\nu^i$ and $\nu^{ci}$ 
together in a lepton triplet, which is described by
\begin{eqnarray}\label{Eq:LeptonTriplets1}
& \psi^i_L = \left( \nu^i, \ell^i, \nu^{ci}\right)^T_L:({\bf 3}, -\frac{1}{3}),  \ \ \ \ell^i_R:({\bf 1}, -1),
\end{eqnarray}
where $Y$ = $-\lambda^8/\sqrt{3}$ + $N$.  
The quark sector consists of two antitriplets with extra heavy down quarks, $d^\prime$, 
and one triplet with an extra heavy up quark, $u^\prime$:
\begin{eqnarray}\label{Eq:QuarkTripletNu}
& Q^{i=1,2}_L  =  \left( d^i,  -u^i,  d^{\prime i}\right)^T_L:({\bf 3^\ast}, 0), \ \ \  
Q^3_L  = \left( u^3,  d^3,  u^{\prime 3}\right)^T_L:({\bf 3}, \frac{1}{3}), 
\nonumber \\
& u^{1,2,3}_R:({\bf 1}, \frac{2}{3}), \ \ \ d^{1,2,3}_R:({\bf 1}, -\frac{1}{3}), \ \ \  
d^{\prime 1,2}_R:  ({\bf 1}, -\frac{1}{3}) , \ \ \ u^{\prime 3}_R:({\bf 1}, \frac{2}{3}).  
\end{eqnarray}
As discussed in Ref.\cite{RadiativeNuDirac}, it is possible to radiatively generate 
neutrino masses of the Dirac type, especially for the electron neutrino, by introducing four Higgs scalars:    
\begin{eqnarray}\label{Eq:HiggsRhoNu}
& \rho = (\rho^+, \rho^0, {\bar \rho^+})^T:({\bf 3}, \frac{2}{3}),  \ \ \
\rho' = (\rho'^+, \rho'^0, {\bar \rho'^+})^T:({\bf 3}, \frac{2}{3}),  
\nonumber \\
& \eta = (\eta^0, \eta^-, {\bar \eta}^0)^T:({\bf 3}, -\frac{1}{3}),  \ \ \
\chi = ({\bar \chi}^0, \chi^-, \chi^0)^T:({\bf 3}, -\frac{1}{3}), 
\end{eqnarray}
and other sextets.  To avoid unwanted flavor changing effects is done by 
the orthogonal choice of  $\langle 0| \eta|0 \rangle$ = 
($v_\eta$, 0, 0)$^T$ and  $\langle 0| \chi|0 \rangle$ = (0, 0, $v_\chi$)$^T$, which is 
ensured by the appropriate Higgs potential\cite{ThreeOptions,DiscrteSymmetry,HiggsPotentialNU}. 
Masses of quarks and leptons are supplied by $\langle 0| \rho^0|0 \rangle$, but the extra Higgs 
scalar, $\rho'$, is requested to develop no VEV, {\it i.e.} $\langle 0| \rho'^0|0 \rangle$ = 0.  

By employing the same Higgs scalars but without sextets, 
one can argue that neutrinos acquire radiatively generated Majorana masses.  
Since $\nu^{ci}_L$ remains massless, its right handed partner, $\nu^{ci}_R$, 
is introduced to form a mass term with $\nu^{ci}_L$.
\footnote{
Instead of introducing $\nu^{ci}_R$, one can use the sextet to develop a Majorana mass term 
of $\nu^{ci}_L\nu^{ci}_L$ by $\langle 0| S^{33}|0 \rangle$ as in Eq.(\ref{Eq:YukawaSextet}).   
However, it needs careful setup for the sextet to develop no Majorana mass of $\nu^i_L\nu^i_L$ by 
$\langle 0| S^{11}|0 \rangle$. 
}
It is the model with heavy neutral leptons ($N^i$)\cite{HeavyNeutral}, where the lepton 
multiplets in Eq.(\ref{Eq:LeptonTriplets1}) are modified to
\begin{eqnarray}\label{Eq:LeptonTriplets2}
& \psi^i_L = \left( \nu^i, \ell^i, N^i\right)^T_L:({\bf 3}, -\frac{1}{3}),  \ \ \ \ell^i_R:({\bf 1}, -1), \ \ \ 
N^i_R:({\bf 1}, 0).
\end{eqnarray}
The Yukawa and Higgs interactions are subject to the following discrete symmetry based on a $Z_2$ parity that 
distinguishes between $\rho$ and $\rho'$ as well as between $\eta$ and $\chi$.  The $Z_2$ parity is negative 
for  ($\psi^i_L$, $Q^{1,2}_L$, $d^{1,2,3}_R$, $u^{\prime 3}_R$, $\rho$, $\chi$) and positive for all others.   

The relevant Higgs coupling, which are responsible for neutrino masses, are specified by
\begin{equation}\label{Eq:HiggsNuRadiative}
\lambda_1\left( \rho^\dagger\eta \right) \left( \chi^\dagger\rho'\right)  + 
\lambda_2\left( \rho^\dagger\rho' \right) \left( \chi^\dagger\eta\right) + ({\rm h.c.}),
\end{equation}
and the Yukawa couplings are given by
\begin{eqnarray}\label{Eq:YukawaNuRight}
-{\cal L}_Y & = &  
\frac{1}{2} \epsilon^{\alpha\beta\gamma} G_{[ij]}{\overline {\left( \psi_{\alpha L}^i \right)^c}}\psi^j_{\beta L}\rho'_\gamma
 +  G_{\ell i}^i{\overline {\psi^i_L}}\ell^i_R \rho
 +  G_{N i}^i{\overline {\psi^i_L}}N^i_R \chi,
\nonumber \\
& + & \sum_{i=1}^{2}\left( {\overline {Q^i_L}}D^i_R \eta^\ast + {\overline {Q^i_L}}U^i_R\rho^\ast
+{\overline {Q^i_L}}C^i_R\chi^\ast \right) 
 +   {\overline {Q^3_L}}U^3_R \eta
\nonumber \\
& + & {\overline {Q^3_L}}D^3_R\rho
 +G_{u^\prime3}{\overline {Q^3_L}}u^{\prime 3}_R\chi + ({\rm h.c.}),
\end{eqnarray}
with $U^i_R$ and $D^i_R$ are the same as in Eq.(\ref{Eq:Yukawa}) and 
$C^i_R$ = $\sum_{j=1}^2G^{\prime i}_{d^\prime j}d^{\prime j}_R$. 
The charged and neutral lepton mass matrices have been set to be diagonal.  The requirement of 
$\langle 0| \rho'^0|0 \rangle$ = 0 should be always maintained to keep neutrinos massless at the 
tree level and is fulfilled as a result of the discrete symmetry.  
The corresponding  one-loop diagrams are given by Fig.1(a) with $\eta^+$ $\rightarrow$ 
${\bar \rho'^+}$ and by Fig.1(b) with ($\chi^-$, $\eta^-$, $\omega^i$) $\rightarrow$ 
(${\bar \chi}^0$, $\rho'^0$, $N^i$).  The resulting neutrino mass matrix is  
reproduced by Eq.(\ref{Eq:NUmass}) with $\delta_{ij}$ =0 and evaluated neutrino masses are 
given by Eq.(\ref{Eq:MatrixEntryIJ}), where the above replacement is performed and 
the coefficient $\lambda$ is replaced by $\lambda_1$ 
($\lambda_2$) for the charged (heavy)-lepton-exchange.  The similar values of $G_{[ij]}$ to the 
previous ones are, thus, obtained.  

The discrete symmetry forbids the Yukawa couplings: 
$\epsilon^{\alpha\beta\gamma}\psi^i_{\alpha L}\psi^j_{\beta L}\rho_\gamma$ and 
${\overline {\psi^i_L}}\ell^i_R \rho'$,  the former of which would yield a mixing term of $\nu^i_LN^j_L$  
at the tree level.  Also forbidden are dangerous Higgs couplings: 
$\epsilon^{\alpha\beta\gamma}\eta_\alpha\rho'_\beta\chi_\gamma$ and 
$\rho^\dagger\rho'$ that would align $\langle 0| \rho'^0|0 \rangle$ $\neq$ 0 and $\eta^\dagger\chi$ that 
would disturb the orthogonal choice of  $\langle 0| \eta|0 \rangle$ and $\langle 0| \chi|0 \rangle$.  
On the other hand, $\epsilon^{\alpha\beta\gamma}\eta_\alpha\rho_\beta\chi_\gamma$ is allowed and gives 
($\eta^0\chi^0$ - ${\bar \eta}^0{\bar \chi}^0$)$\rho^0$ in the Higgs potential.  
The VEV's of $\langle 0| \eta^0|0 \rangle$ and $\langle 0| \chi^0|0 \rangle$ are preferred by  
$\eta^0\chi^0\rho^0$ giving a negative contribution to the Higgs potential, which corresponds to  
a positive contribution of ${\bar \eta}^0{\bar \chi}^0\rho^0$. Therefore,  
$\epsilon^{\alpha\beta\gamma}\eta_\alpha\rho_\beta\chi_\gamma$ helps to align 
$\langle 0| {\bar \eta}^0|0 \rangle$ = 0 and $\langle 0| {\bar \chi}^0|0 \rangle$ = 0.  
The non-self Hermitian part of the Higgs potential includes 
$\epsilon^{\alpha\beta\gamma}\eta_\alpha\rho_\beta\chi_\gamma$, 
$(\chi^\dagger\eta)^2$, $(\rho^\dagger\rho^\prime)^2$ and Eq.(\ref{Eq:HiggsNuRadiative}).  
Finally, since the $SU(2)_L$-singlet exotic quarks have the same charges as the $SU(2)_L$-doublet 
ordinary  quarks, there are dangerous flavor-changing Higgs interactions\cite{FlavorChangingGeneral}. 
For example, they can be suppressed by restricting the Yukawa interactions for quarks by imposing  
the discrete symmetry. The corresponding $Z_2$ parity is negative for  
($\psi^i_L$, $u^{1,2}_R$, $u^{\prime 3}_R$, $d^3_R$, $d^{\prime 1,2}_R$, $\rho$, $\chi$)\cite{GIMinSU3} 
instead of the previous one:  ($\psi^i_L$, $Q^{1,2}_L$, $d^{1,2,3}_R$, $u^{\prime 3}_R$, $\rho$, $\chi$).

It should be noted that, in the present context with the assignment of ($\nu^i$, $\ell^i$, $\nu^{ci}$) 
as in Eq.(\ref{Eq:LeptonTriplets1}), 
Dirac neutrino masses would be generated if there were 
the transition of ${\bar \rho'^+}$ $\leftrightarrow$ ${\bar \rho}^+$.  It is possible by the Higgs couplings of 
$\left( \rho^\dagger\chi \right) \left( \chi^\dagger\rho' \right)$, which would not disturb 
$\langle 0| \rho'^0|0 \rangle$ = 0.  However, if this coupling is allowed, one cannot  
forbid $\rho^\dagger\rho'$ by any symmetries, which necessarily causes 
$\langle 0| \rho'^0|0 \rangle$ $\neq$ 0.  Then, Dirac neutrino masses are present at the tree level.

In summary, we have succeeded in accommodating a radiative mechanism for Majorana neutrino masses 
in models based on $SU(3)_L$ $\times$ $U(1)_N$. Heavy leptons are required for the mechanism to 
consistently work.   In the model with heavy neutral leptons, it is essential to employ one 
$\ddash$silent" Higgs scalar that is nothing to do with the symmetry breakdown. On the other hand, 
the model including heavy charged leptons needs no extra Higgs scalars.  All Higgs scalars are just 
contained in a minimal set, which gives a correct symmetry breaking of 
$SU(3)_L$ $\times$ $U(1)_N$ as well as masses of quarks and leptons.  
The essence lies in the property that a Higgs triplet necessarily includes a charged Higgs scalar that 
couples to a charged lepton-neutrino pair, which plays an important role in the radiative mechanism. 
We hope that one of the models based on $SU(3)_L$ $\times$ $U(1)_N$ serves as the promising model 
beyond the standard model as far as radiative generation of neutrino masses is concerned.


\noindent
{\bf \center Figure Caption\\}
	
\noindent
\begin{description}
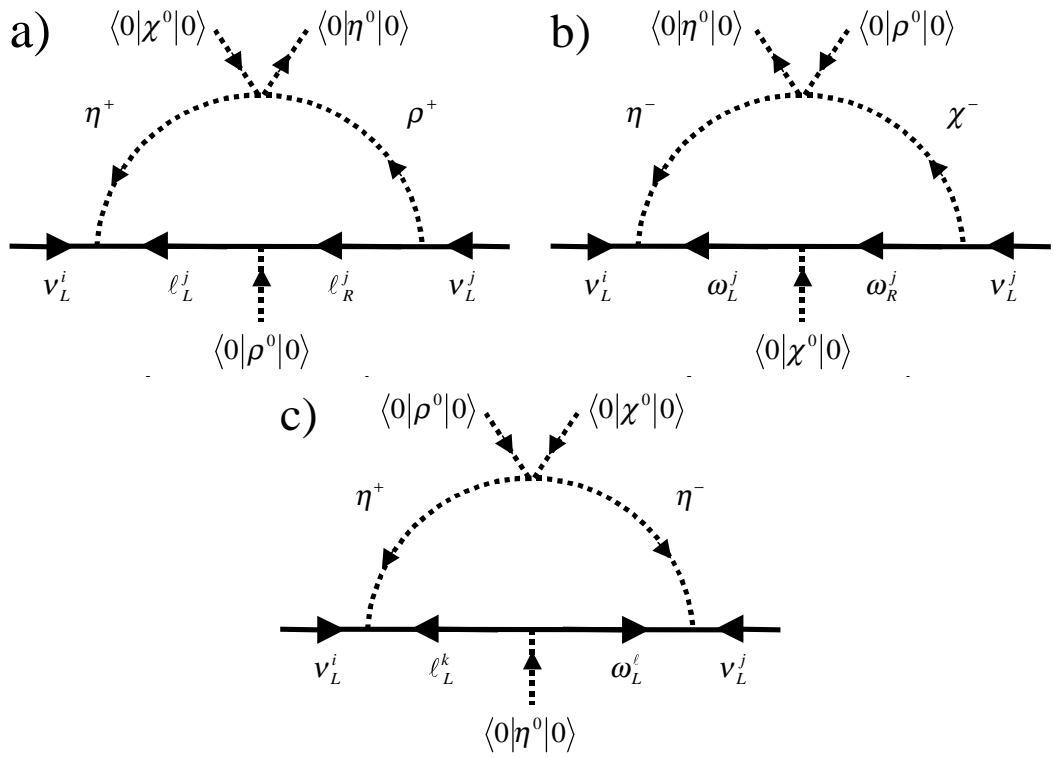
\item {\bf Fig.1} : One loop radiative diagrams for $\nu^i$-$\nu^j$ via (a) charged leptons and (b) heavy leptons for $i\neq j$
 and via (c) the mixing between charge leptons and heavy leptons for $i\neq k$, $k\neq \ell$ and 
$\ell\neq j$.
\end{description}


\begin{references}
%
\bibitem{EarlyMassive} Z. Maki, M. Nakagawa and S. Sakata, Prog. Theor. Phys. {\bf 28} (1962) 870. 
See also  B. Pontecorvo, JETP (USSR) {\bf 34} (1958) 247;  
B. Pontecorvo, Zh. Eksp. Teor. Piz. {\bf 53} (1967) 1717;
V. Gribov and B. Pontecorvo, Phys. Lett. {\bf 28B} (1969) 493. 

\bibitem{SuperKamiokande} SuperKamiokande Collaboration, Y. Fukuda {\it et al.}, 
Phys.  Rev. Lett. {\bf 81} (1998) 1562; Phys. Lett. B {\bf 433} (1998) 9 and {\bf 436} (1998) 33. 
See also K. Scholberg, hep-ex/9905016 (May, 1999).
%

\bibitem{SolarDeficit} See for example, J.N. Bahcall, P.I. Krastev and A.Yu. Smirnov, 
Phys. Rev. D {\bf 58} (1998) 096016-1; J.N. Bahcall, astro-ph/9808162 (Aug., 1998).

%
\bibitem{LSND} C. Athanassopoulos, Phys. Rev. Lett. {\bf 75} (1995) 2650; 
{\bf 77} (1996) 3082; {\bf 81} (1998) 1774. 

%
\bibitem{LaboNu} A.O. Bazarko, hep-ex/9906003 (June, 1999) and other references therein.

%
\bibitem{SeeSaw} T. Yanagida, in {\it Proceedings of the Workshop on Unified  
Theories and Baryon Number in the Universe} edited by A. Sawada and A. Sugamoto 
(KEK Report No.79-18, Tsukuba, 1979), p.95; Prog. Theor. Phys. {\bf 64} (1980) 1103;  
M. Gell-Mann, P. Ramond and R. Slansky, in {\it Supergravity} edited by P. van 
Nieuwenhuizen and D.Z. Freedmann (North-Holland, Amsterdam 1979), p.315; 
R.N. Mohapatra and G. Senjanovic, Phys. Rev. Lett. {\bf 44} (1980) 912.

%
\bibitem{Radiative} A. Zee, Phys. Lett. {\bf 93B} (1980) 389; {\bf 161B} (1985) 141. 
See also L. Wolfenstein, Nucl. Phys. {\bf B175} (1980) 93; 
S. P. Petcov, Phys. Lett. {\bf 115B} (1982) 401;    
K. S. Babu and V. S. Mathur, Phys.  Lett. B {\bf 196} (1987) 218;   
K. S. Babu, Phys.  Lett. B {\bf 203} (1988) 132;   
J. Liu, Phys.  Lett. B {\bf 216} (1989) 367;    
W. Grimus and H. Neufeld, Phys.  Lett. B {\bf 237} (1990) 521;    
B. K. Pal, Phys. Rev. D {\bf 44} (1991) 2261;  
W. Grimus and G. Nardulli, Phys.  Lett. B {\bf 271} (1991) 161;    
A. Yu. Smirnov and Z. Tao, Nucl. Phys. B {\bf 426} (1994) 415.

%
\bibitem{SU3U1} 
F. Pisano and V. Pleitez, Phys. Rev. D {\bf 46} (1992) 410; 
P.H. Frampton, Phys. Rev. Lett. {\bf 69} (1992) 2889; 
V. Pleitez and M.D. Tonasse, Phys. Rev. D {\bf 48} (1993) 2353;
D. Ng,   Phys. Rev. D {\bf 49} (1994) 4805;
M. $\ddot{{\rm O}}$zer, Phys. Rev. D {\bf 54} (1996) 1143.  
For earlier works, see for example, J. Schechter and Y. Ueda, Phys. Rev. D {\bf 8} (1973) 484; 
H. Fritzsch and P. Minkowski, Phys. Lett. {\bf 63B} (1976) 99; 
M. Singer, J.M.F. Alle and J. Schechter, Phys. Rev. D {\bf 22} (1980) 738.

%
\bibitem{RhandedNU} 
R. Foot, H.N. Long and T.A. Tran, Phys. Rev. D {\bf 50} (1994) 34; 
H.N. Long and T.A. Tran, Mod. Phys. Lett. {\bf A9} (1994) 2507;  
H.N. Long, Phys. Rev. D {\bf 54} (1996) 4691;  
F.-z. Chen, Phys. Lett. B {\bf 442} (1998) 223. 
H.N. Long and T. Inami, hep-ph/9902475 (Feb., 1999). 

%
\bibitem{Sextet} R. Foot, O.F. Hern$\acute{{\rm a}}$ndez, F. Pisano and V. Pleitez, Phys. Rev. D {\bf 47} (1993) 4158; 
M.D. Tonasse,  Phys. Lett. B {\bf 381} (1996) 191; 
H.N. Long, Mod. Phys. Lett. {\bf A13} (1998) 1865; 
N.T. Anh, N.A. Ky and H.N. Long, hep-ph/9810273 (Oct., 1998); 
M.B. Tully and G.C. Joshi, hep-ph/9810282 (Oct., 1998).

%
\bibitem{RadiativeNuDirac} R. Barbieri and R.N. Mohapatra, Phys. Lett. B {\bf 218} (1989) 225; 
J. Liu, Phys. Lett. B {\bf 225} (1989) 148; 

%
\bibitem{RadiativeNuSU3} V. Pleitez and M.D. Tonasse, Phys. Rev. D {\bf 48} (1993) 5274; 
F. Pisano, V. Pleitez and M.D. Tonasse, hep-ph/9310230 v2 (Feb., 1994); 
P.H. Frampton, P.I. Krastev and J.T. Liu, Mod. Phys. Lett. {\bf A9} (1994) 761; 

%
\bibitem{MajoranaNuSextet}
M. $\ddot{{\rm O}}$zer, Phys. Lett. B {\bf 337} (1994) 324; 
F. Pisano, J.A. Silva-Sobrinho and M.D. Tonasse, Phys. Lett. B {\bf 388} (1996) 338; 
V. Pleitez and M.D. Tonasse, Phys. Lett. B {\bf 430} (1998) 174.

%
\bibitem{GIMinSU3} 
C. Montero, F. Pisano and V. Pleitez, Phys. Rev. D {\bf 47} (1993) 2918; 
M. $\ddot{{\rm O}}$zer, Phys. Rev. D {\bf 54} (1996) 4561; 

%
\bibitem{SUSYSU3U1} T.V. Duong and E. Ma, Phys. Lett. B {\bf 316} (1994) 307; 
H.N. Long and P.B. Pal, Mod. Phys. Lett. {\bf A13} (1998) 2355.

%
\bibitem{SU3U1Application} 
J.T. Liu, Phys. Lett. B {\bf 225} (1989) 148; Phys. Rev. D {\bf 50} (1994) 542; 
J.T. Liu and D. Ng, Phys. Rev. D {\bf 50} (1994) 548; Zeit Phys. {\bf C62} (1994) 693; 
P.H. Frampton, J.T. Liu, B.C. Rasco and D. Ng, Mod. Phys. Lett. {\bf A9} (1994) 1975; 
C.-s. Huang and T.-j. Li, Phys. Rev. D {\bf 50} (1994) 2127; Zeit Phys. {\bf C68} (1995) 319;   
D. G. Dumm, F. Pisano and V. Pleitez, Mod. Phys. Lett. {\bf A9} (1994) 1609; 
R. Foot, Mod. Phys. Lett. {\bf A10} (1995) 159; 
L. Epele, H. Fanchiotti, C.G. Canal and D.G. Dumm, Phys. Lett. B {\bf 343} (1995) 291; 
P.B. Pal, Phys. Rev. D {\bf 52} (1995) 1659; 
V. Pleitez, Phys. Rev. D {\bf 53} (1996) 514; 
D.G. Dumm, Int. J. Mod. Phys. {\bf A11} (1996) 887; Phys. Lett. B {\bf 411} (1997) 313; 
J.C. Montero, V. Pleitez and M.C. Rodriguez, Phys. Rev. D {\bf 58} (1998) 097505;  
P. Das and P. Jain, hep-ph/9903432 (Mar., 1999); 
M.B. Tully and G.C. Joshi, hep-ph/9905552 (May, 1999).

%
\bibitem{SU3U1Muonium} 
H. Fujii, Y. Mimura, K. Sasaki and T. Sasaki, Phys. Rev. D {\bf 49} (1994) 559;
P.W. Ko, Nuovo Cim. {\bf A107} (1994) 809; 
V. Pleitez, hep-ph/9905406 (May, 1999).

%
\bibitem{MassNu} See for example, A.Yu. Smirnov,  hep-ph/9901208 (Jan., 1999).

%
\bibitem{ThreeOptions} C. Montero, F. Pisano and V. Pleitez, in Ref.\cite{GIMinSU3}

%
\bibitem{DiscrteSymmetry}  R. Foot, O.F. Hern$\acute{{\rm a}}$ndez, F. Pisano and V. Pleitez, in Ref.\cite{Sextet}.

%
\bibitem{HeavyLepton} V. Pleitez and M.D. Tonasse, in Ref.\cite{SU3U1}. 

%
\bibitem{FineTuneNu} 
V. Pleitez and M.D. Tonasse, in Ref.\cite{RadiativeNuSU3}; 
P.H. Frampton, P.I. Krastev and J.T. Liu, in Ref.\cite{RadiativeNuSU3}; 
M.B. Tully and G.C. Joshi, in Ref.\cite{Sextet}. 

%
\bibitem{DangerousCoupling} L. Epele, H. Fanchiotti, C.G. Canal and D.G. Dumm, in Ref.\cite{SU3U1Application}; 
F. Pisano, J.A. Silva-Sobrinho and M.D. Tonasse, in Ref.\cite{MajoranaNuSextet}.

%
\bibitem{Ma}  E. Ma, Phys. Rev. Lett. {\bf 81} (1998) 1171.

%
\bibitem{ZeeModelNU}  C. Jarlskog, M. Matsuda, S. Skadhauge and M. Tanimoto, 
 Phys. Lett. B {\bf 449} (1999) 241. See also, A.Yu. Smirnov and M. Tanimoto, 
Phys. Rev. D {\bf 55} (1997) 1665.

%
\bibitem{HiggsPotentialNU}  P.B. Pal, in Ref.\cite{SU3U1Application}; 
H.N. Long, in Ref.\cite{Sextet}.

%
\bibitem{HeavyNeutral} F.-z. Chen, in Ref.\cite{RhandedNU}.  
See also V. Pleitez and M.D. Tonasse, in Ref.\cite{MajoranaNuSextet}. 

%
\bibitem{FlavorChangingGeneral}  
S.L. Glashow and S. Weinberg, Phys. Rev. D {\bf 15} (1977) 1985; 
H. Georgi and A. Pais, Phys. Rev. D {\bf 19} (1979) 2746.
\vspace{10mm}
\end{references}
\end{document}